\documentclass{article}
\usepackage{arxiv}
\usepackage{natbib}
\usepackage[utf8]{inputenc}
\usepackage{amsmath,amssymb,amsfonts}
\usepackage{algorithmic}
\usepackage{graphicx}
\usepackage{textcomp}
\usepackage{xcolor}
\usepackage{booktabs}
\usepackage{multirow}
\usepackage{longtable}
\usepackage{paralist}
\usepackage{hyperref}
\usepackage{array}
\usepackage{tabulary}
\usepackage{makecell}
\usepackage{cellspace}
\usepackage{longtable}
\usepackage[acronym,nonumberlist,nogroupskip,noredefwarn]{glossaries}
\newcolumntype{L}[1]{>{\raggedright\let\newline\\\arraybackslash\hspace{0pt}}m{#1}}
\newcolumntype{C}[1]{>{\centering\let\newline\\\arraybackslash\hspace{0pt}}m{#1}}
\newcolumntype{R}[1]{>{\raggedleft\let\newline\\\arraybackslash\hspace{0pt}}m{#1}}

\hypersetup{
  colorlinks,
  citecolor=black,
  filecolor=black,
  linkcolor=black,
  urlcolor=black,
  pdfauthor={Eduardo Freitas},
  pdftitle={},
  pdfkeywords={},
  pdfsubject={},
  pdfcreator={GNU Emacs with AUCTeX Package},
  pdflang={en}
}

\newacronym{vmm}{VMM}{Virtual Machine Monitor}
\newacronym{vfs}{VFS}{Virtual File System}
\newacronym{ufpe}{UFPE}{Universidade Federal de Pernambuco}
\newacronym{nic}{NIC}{Network Interface Card}
\newacronym{io}{I/O}{Input/Output}
\newacronym{os}{OS}{Operating System}
\newacronym{vm}{VM}{Virtual Machine}
\newacronym{fifo}{FIFO}{First-In, First-Out}
\newacronym{kvm}{KVM}{Kernel-based Virtual Machine}
\newacronym{ddos}{DDoS}{Distributed Denial-of-Service}
\newacronym{sdn}{SDN}{Software Defined Networks}
\newacronym{iot}{IoT}{Internet of Things}
\newacronym{it}{IT}{Information Technology}
\newacronym{nasa}{NASA}{National Aeronautics and Space Administration}
\newacronym{dt}{DT}{Digital Twin}
\newacronym{ndt}{NDT}{Network Digital Twin}
\newacronym{dtn}{DTN}{Digital Twin Network}
\newacronym{plm}{PLM}{Product Lifecycle Management}
\newacronym{MQTT}{MQTT}{Message Queuing Telemetry Transport}
\newacronym{AMQP}{AMQP}{Advanced Message Queuing Protocol}
\newacronym{DDS}{DDS}{Data Distribution Service}
\newacronym{rpc}{RPC}{Remote Procedure Call}
\newacronym{ipfix}{IPFIX}{IP Flow Information Export}
\newacronym{ai}{AI}{Artificial Inteligence}
\newacronym{cps}{CPS}{Cyber-Physical Systems}
\newacronym{dtaas}{DTaaS}{Digital Twins as a Service}
\newacronym{ml}{ML}{Machine Learning}

\title{Digital Twin Synchronization: towards a data-centric architecture\thanks{This is a preprint. The final published version may contain differences.}}

\author{%
    \\
    \textbf{Eduardo Freitas} \\
    Centro de Informática (CIn) \\
    Grupo de Pesquisa em Redes e Telecomunicações\\(GPRT)\\
    Universidade Federal de Pernambuco (UFPE) \\
    Recife, Brasil \\
    \texttt{eduardo.freitas@gprt.ufpe.br}
    \and
    \\
    \textbf{Assis T. de Oliveira Filho} \\
    Centro de Informática (CIn) \\
    Grupo de Pesquisa em Redes e Telecomunicações\\(GPRT)\\
    Universidade Federal de Pernambuco (UFPE) \\
    Recife, Brasil \\
    \texttt{assis.tiago@gprt.ufpe.br}
    \and
     \\
    \textbf{Pedro R. X. Carmo} \\
    Centro de Informática (CIn) \\
    Grupo de Pesquisa em Redes e Telecomunicações\\(GPRT)\\
    Universidade Federal de Pernambuco (UFPE) \\
    Recife, Brasil \\
    \texttt{pedro.carmo@gprt.ufpe.br}
    \and
    \\
    \textbf{Djamel Sadok} \\
    Centro de Informática (CIn) \\
    Grupo de Pesquisa em Redes e Telecomunicações\\(GPRT)\\
    Universidade Federal de Pernambuco (UFPE) \\
    Recife, Brasil \\
    \texttt{jamel@gprt.ufpe.br}
    \and
    \\
    \textbf{Judith Kelner} \\
    Centro de Informática (CIn) \\
    Grupo de Pesquisa em Redes e Telecomunicações\\(GPRT)\\
    Universidade Federal de Pernambuco (UFPE) \\
    Recife, Brasil \\
    \texttt{jk@gprt.ufpe.br}
}
\renewcommand{\shorttitle}{}

\hypersetup{
pdftitle={Digital Twin Synchronization: towards a data-centric architecture},
pdfsubject={cs.NI},
pdfauthor={Eduardo Freitas}
}

\begin{document}

\maketitle

\begin{abstract}
  Digital Twin (DT) technology revolutionizes industrial processes by enabling the representation of physical entities and their dynamics to enhance productivity and operational efficiency. It has emerged as a vital enabling technology in the Industry 4.0 context. The present article examines the particular issue of synchronizing a digital twin while ensuring an accurate reflection of its physical counterpart. Despite the reported recent advances in the design of middleware and low delay communication technologies, effective synchronization between both worlds remains challenging. This paper reviews currently adopted synchronization technologies and architectures, identifies vital outstanding technical challenges, and proposes a unified synchronization architecture for use by various industrial applications while addressing security and interoperability requirements. As such, this study aims to bridges gaps and advance robust synchronization in DT environments, emphasizing the need for a standardized architecture to ensure seamless operation and continuous improvement of industrial systems.
\end{abstract}


\section{Introduction}
\label{sec:introduction}
Industry 4.0 strives for the advanced integration of digital technologies into manufacturing processes and industrial operations. The primary enabler behind this tendency is the development and adoption of emerging technologies such as the \gls{iot}, Big Data, Cloud Computing, \gls{ai}, and \gls{cps}. Industry 4.0 seeks to create intelligent factories where machines, systems, and humans communicate and collaborate continuously and autonomously toward common production goals. Such a highly digitalized environment is expected to support the optimization of production processes, a mass adaptation of products, greater flexibility and operational efficiency, and the ability to respond quickly to changing market demands.

By creating precise virtual replicas of physical entities, the \gls{dt} technology facilitates the seamless monitoring, simulation, and optimization of industrial processes, aligning itself perfectly with the goals of Industry 4.0. It offers a way to enhance productivity, flexibility, and operational efficiency. A critical factor in this technology is the effective synchronization between the virtual and physical worlds. Synchronization is especially vital when implementing \glspl{dt} for industrial environments \cite{8477101}. It enables near real-time bidirectional updating, where data collected from physical sensors and devices is transferred to the virtual model, influencing future decisions over the physical environment.

Despite advances in digital twin technology and middleware efficiency, synchronization still faces challenges. For instance, the diversity of encountered devices causes significant data heterogeneity, which makes synchronization a complex process \cite{Liu2020Review}. In addition, the considerable number of industry devices is susceptible to generating a massive amount of data, posing challenges concerning their processing and storage. This requires the assistance of special technologies like Big Data and \gls{ai} \cite{JEREMIAH2024103120}, which further emphasizes the importance of a \textit{data-centric} synchronization architecture.

Throughout the existing literature, there are multiple studies detailing approaches toward the support of synchronization. However, most of these focus on particular use cases, highlighting the need for a general approach with broader applicability. The goal is to design a synchronization architecture that can easily be tailored to fit multiple industrial scenarios. Indeed, several studies underscore the importance of a unified architecture. For instance, Bellavista \cite{Bellavista2021Application-Driven} emphasizes the necessity of such an architecture to manage network heterogeneity and enhance DT management. Similarly, Kherbache \cite{Kherbache2022Network} highlights the need for a cohesive architecture to support diverse industrial applications of digital twins. Given the diversity of synchronization approaches in the literature, we realized the need for a study highlighting the leading works in the field and, most importantly, to develop a unified reference architecture.

The main goal of this research is to propose a Data-Centric Industrial Digital Twin Synchronization Architecture that can be sufficiently generic enough to embrace the diversity of Industry 4.0 application scenarios. We provide an overview of the technologies, challenges, and opportunities associated with synchronizing a Digital Twin and the real world, exploring existing literature, standards, and tools that can provide such architecture.

This paper is structured as follows: Section \ref{sec:dt} discusses the concept of digital twins and synchronization and provides a literature review of \glspl{dt} in the context of Industry 4.0. Section \ref{sec:dts-design} explores the design of the digital twin synchronization architecture, including critical components and their functions. Section \ref{sec:telemetry} focuses on Telemetry, detailing data collection, transformation, and storage. Section \ref{sec:analysis} presents the analysis of update events. Finally, Section \ref{sec:conclusion} offers conclusions and future directions for research on digital twin synchronization.

\section{Digital Twin}
\label{sec:dt}
Though generating a great deal of interest recently, the concept of the Digital Twin is not entirely new. As acknowledged by previous papers \cite{Liu2020Review, 8477101, Grieves2017}, it was introduced for the first time in 2002 as a Product Lifecycle Management tool. In this concept, a ``virtual space'' would mirror the ``real space'', and data would flow between these spaces to enable proper synchronization. Only in the year 2010, \gls{nasa} published a technical report offering the first known accurate definition of Digital Twin:  ``\textit{A digital twin is an integrated multiphysics, multiscale simulation of a vehicle or system that uses the best available physical models, sensor updates, fleet history, etc., to mirror the life of its corresponding flying twin. The digital twin is ultra-realistic and may consider one or more important and interdependent vehicle systems [...]}'' \cite{nasa-draft}. Although initially focused on aeronautics, this definition broadened the concept's applicability to multiple fields. Following this, the white paper \cite{grieves} was published in 2014, providing the first concept model and use cases to the literature. The paper enabled the idea to gain attention and practical implementations, especially in the Industry field.

Despite the growing interest in Digital Twin across the literature, there is yet no unique formal definition for it \cite{Liu2020Review}. Multiple published works in the literature adopt different conceptual models and particular interpretations of existing definitions. This is often dependent on the adopted scenario requirements. Nonetheless, it remains possible to identify a common viewpoint for a Digital Twin where it is seen as a \textit{realistic virtual replication of real-life entities that is constantly being updated by this real entity and also continually updates the real entity}. This idea explains the basic concepts of a three-part digital twin architecture: physical entity, digital entity, and the data flow between them. 

A physical entity typically needs little attention from the DT concept viewpoint, given that it represents an actual industry device, such as an \gls{iot} device, a robot, a sensor, a network router, or a camera. On the other hand, there are multiple ways to build the digital entity, including the use of mathematical models to simulate physical behavior \cite{9865226, 10070572, Cheng2023Digital-Twin-Based}, or complete three-dimensional emulation of physical entities inside 3D frameworks \cite{10116025, exploiting-cell}.

As for the third part, the data flow between digital and physical entities, research has highlighted the importance of bidirectional updates for the DT \cite{Liu2020Review}, which we refer to in this paper as \textit{Digital Twin Synchronization}. \gls{dt} synchronization can be achieved in multiple ways, similarly to the digital entity itself. Mathematical models \cite{Gehrmann2020A, 9865226, LEUNG2022108353}, process simulation \cite{Jia2023Accurate, SCHNURER20222416}, or middleware communication \cite{CSI, zcz-nmrg-digitaltwin-data-collection-03} are candidates. However, as we will see in this section, many of these options are unsuitable for a real-world industry scenario.

\subsection{Digital Twin and Industry}
\label{sec:dt-industry}
Currently, \gls{dt}'s applicability is mainly focused on the Industry field \cite{9103025}. They have been leveraged across multiple Industry 4.0 domains, including smart manufacturing, predictive maintenance, and automated logistics, enhancing operational efficiency and process quality \cite{10188847}. Given their success, they are also explored in network, healthcare, construction, agriculture, and urban management areas. For instance, in healthcare, patients may use digital twins to plan surgeries and personalized treatment \cite{Pires2019Digital}. In the construction domain, they may assist in infrastructure monitoring and maintenance \cite{Huang2021etal}. In urban management, one can create digital models of entire cities, aiding in resource planning and management \cite{Pires2019Digital, Schroeder2021A}.

Special attention is necessary toward the concept of \gls{ndt}. It is a special domain of the Digital Twin concept, focusing specifically on the modeling, simulation, and optimization of communication networks. As network size and complexity increase, the importance of Digital Twins in network optimization and development becomes increasingly evident \cite{10175444, 9429703}. The concept of DT in communication networks has led to the development of \glspl{ndt}, replicating network infrastructure complexity in virtual models reflecting networks' current configuration and operational state. NDTs allow experimentation with configuration changes, impact analysis of different scenarios, and proactive identification of potential failures and bottlenecks \cite{zhou-nmrg-digitaltwin-network-concepts-07}.

Technologies like virtualization have existed for a long time \cite{FILHO202273}. Virtual machines can be considered a primitive form of Network Digital Twin, as they emulate the behavior of real servers and can interact with other VMs as real servers. Although virtualization alone does not constitute a complete Digital Twin, it can facilitate a complete NDT where other technologies can promote proper synchronization and interaction with real devices. However, this approach offers challenges, particularly in computational efficiency \cite{9796944}. One of the most critical motivations for NDT is the ability to create large networks with complex configurations. However, there are better tools than virtualization, as each VM can consume substantial resources to emulate real network devices. Simulation has been a promising approach for creating virtual NDT models \cite{zhou-nmrg-digitaltwin-network-concepts-07}. Recent emphasis has been on Machine Learning models, particularly Graph Neural Networks (GNNs) \cite{9310275, 9795043, Mozo_2022}. Other techniques, such as deep reinforcement learning and knowledge graphs, have also been identified as promising.

At the heart of Industry 4.0, \textit{Smart Manufacturing} \cite{Klaus-Dieter} brings the use of communication and information technology to enable seamless integration between \gls{iot} devices, robots, and human workers. Smart Manufacturing seeks new ways to integrate technologies like Cyber-Physical Systems \cite{NEGRI2017939}, Big Data \cite{10.1063/1.5031520}, and Cloud Computing. The next step in this integration is combining Digital Twin technology with smart manufacturing \cite{Grieves2017, LIU2021etal}. 

Ali et al. \cite{Ali2021etal} describe how Digital Twins are used in smart manufacturing to monitor, analyze, and simulate physical assets and processes. They highlight that current applications focus on status monitoring, simulation, and visualization, allowing real-time analysis of equipment and process conditions and contributing to production optimization. The study uniquely emphasizes real-time visualization and simulation.

For example, in recent years, Siemens has implemented Digital Twins in its Amberg Electronics Plant to simulate and optimize production processes. This innovation has yielded promising results, including a 1400\% increase in productivity \cite{siemens2019}. Academic research has further explored the potential and challenges of Digital Twins in manufacturing. In the work by Wang et al. \cite{Wang2021etal}, a comprehensive review of existing literature on Digital Twins is presented, discussing their evolution, applications, definitions, and challenges. This study provides a perspective on future research needs and opportunities for Digital Twins, particularly semiconductor manufacturing.

Similarly, Onaji et al. \cite{Onaji2022etal} investigate the development and significance of Digital Twins in smart manufacturing, presenting a conceptual framework that supports product-process integration. Through case studies, they demonstrate the benefits of Digital Twins in improving manufacturing efficiency and quality. These academic contributions highlight the growing importance of Digital Twins in industrial practice.

\textit{Predictive maintenance} represents another significant application of Digital Twins in the Industry domain. Aheleroff et al. \cite{aheleroff2021etal} propose a reference architecture model for Digital Twins as a Service (DTaaS), applying it to diverse industrial cases, including predictive maintenance. Their study highlights using Digital Twins for real-time monitoring and predictive maintenance, improving operational efficiency.

Some companies have, for some time, realized the benefits of developing Digital Twin software. A case in point is GE Aviation. It leverages Digital Twins to predict maintenance needs for its jet engines. The company announced around a 20\% reduction in maintenance costs, a 75\% reduction in time to achieve outcomes, and \$11 million in avoided lost production achieved by detecting and preventing three failures \cite{gedigitaltwin}. Lattanzi et al. \cite{Lattanzi2021etal} analyze related aspects of monitoring and predictive maintenance capabilities. Their paper discusses how digital systems can create virtual copies of the physical world to optimize and control manufacturing processes. Furthermore, Mortlock et al. \cite{Mortlock2022etal} present cognitive Digital Twins as the next evolution of this technology, enabling autonomous decisions and continuous improvement in manufacturing.

The application of Digital Twins in optimizing \textit{power distribution networks} has become a crucial innovation for efficient and resilient energy management. For example, ABB has enhanced the efficiency and reliability of power distribution systems. In a specific case study, ABB managed a 15\% reduction in energy losses and improved system reliability, resulting in a 10\% annual reduction in operational costs \cite{abb_digital_twin}.

Mylonas et al. \cite{Mylonas2021etal} review the application of Digital Twins in smart cities and Industry 4.0, emphasizing their importance in energy management. The study discusses the challenges and solutions for implementing Digital Twins in power networks. Mihai et al. \cite{Mihai2022etal} provides a comprehensive study on how Digital Twin can promote electricity distribution network optimization, predicting demands and preventing overloads. Jiang et al. \cite{Jiang2021etal} discusses the capabilities of Digital Twins to monitor and optimize industrial plants' energy efficiency and operational performance.

Regarding the \textit{automotive manufacturing} domain, Digital Twins have been applied to optimize the routes and schedules of Automated Guided Vehicles (AGVs), reducing downtime and improving overall productivity \cite{Martinez-Gutierrez2021etal}. The works by Martinez et al. \cite{Martinez2021etal} and Lichtenstern and Kerber \cite{Lichtenstern2022etal} offer insights into the application of digital twins in improving the performance of AGVs in a production environment. Martinez et al. discuss a digital twin model that simulates the performance of AGVs and improves the accuracy of production planning. The approach uses a web simulation service that allows users to interact and visualize AGV operations in real time, optimizing transportation processes in manufacturing. Lichtenstern and Kerber develop a digital two-fold model based on data reflecting the behavior of AGVs and transport order management. This model replicates vehicle behavior, optimizes fleet size, and evaluates different network topologies, increasing productivity in hybrid flow pre-assembly. Both studies highlight the potential of digital twins to improve efficiency, accuracy, and user interaction in industrial transportation systems.

The application of digital twins in Industry is well illustrated by a case study from a German multinational company, where digital twins were used to optimize production processes and improve efficiency. This study revealed that digital twins not only improve the accuracy of task planning and execution but are also essential for the agility and adaptability of processes. Successful implementations involved the balanced integration of technological, human, and process factors, delivering significant incremental improvements in productivity and operational efficiency \cite{german7x1brasil}.

\subsection{Digital Twin Synchronization}
\label{sec:dt-sync}
The previous section demonstrates, through the numerous studies, the relevance of DTs and their effectiveness in industry across diverse case studies. Also, the synchronization of the two worlds emerges as a factor that determines success \cite{JEREMIAH2024103120, JIANG2022104397}. Synchronization ensures that the virtual models inside the DT produce results based on the most up-to-date and realistic data \cite{Zipper2019Synchronization, Akbarian2020Synchronization}. It is also responsible for translating the knowledge derived from the DT models to the real world, offering benefits such as error prevention and optimization. Synchronization can be achieved by combining techniques from multiple fields of expertise, such as Big Data techniques, Network Telemetry, Cloud Computing, and Artificial Intelligence. Given the wide scope nature of the adopted mechanisms, synchronization warrants a dedicated examination.

Typically, an industry scenario consists of \gls{iot} devices, robotic arms, mobile robots, humans, cameras, network devices, sensors, and specialized instruments, among other objects. Overall, creating a DT using solely pure modeling and/or simulation is insufficient to embrace every aspect of such a complex scenario. On the other hand, modeling each device aspect can be exhausting, as can predicting its behavior. For such a complex scenario, other tools can be combined instead. They include 3D robotics simulators, network device virtualization, or hardware emulation \cite{10.22260/ISARC2020/0205, tahemaa2019digital}. These tools can create broader digital versions of specific industry devices, hence considering the specific characteristics of each one.

Considering such tools, synchronization becomes even more complex than expected since the level of DT device and component heterogeneity is higher. The use of new virtual devices with specific digital technologies results in a higher diversity of data, which, in turn, leads to higher data collection and update executions. This fact yields all sorts of complexity inside DT Synchronization, creating the necessity for a convoluted architecture that can accommodate the complexity of the industry scenario, the DT scenario, and the need to synchronize the two systems.

\section{Digital Twin Synchronization Design}
\label{sec:dts-design}
\begin{figure}[ht]
  \centering
  \includegraphics[width=0.4\textwidth]
  {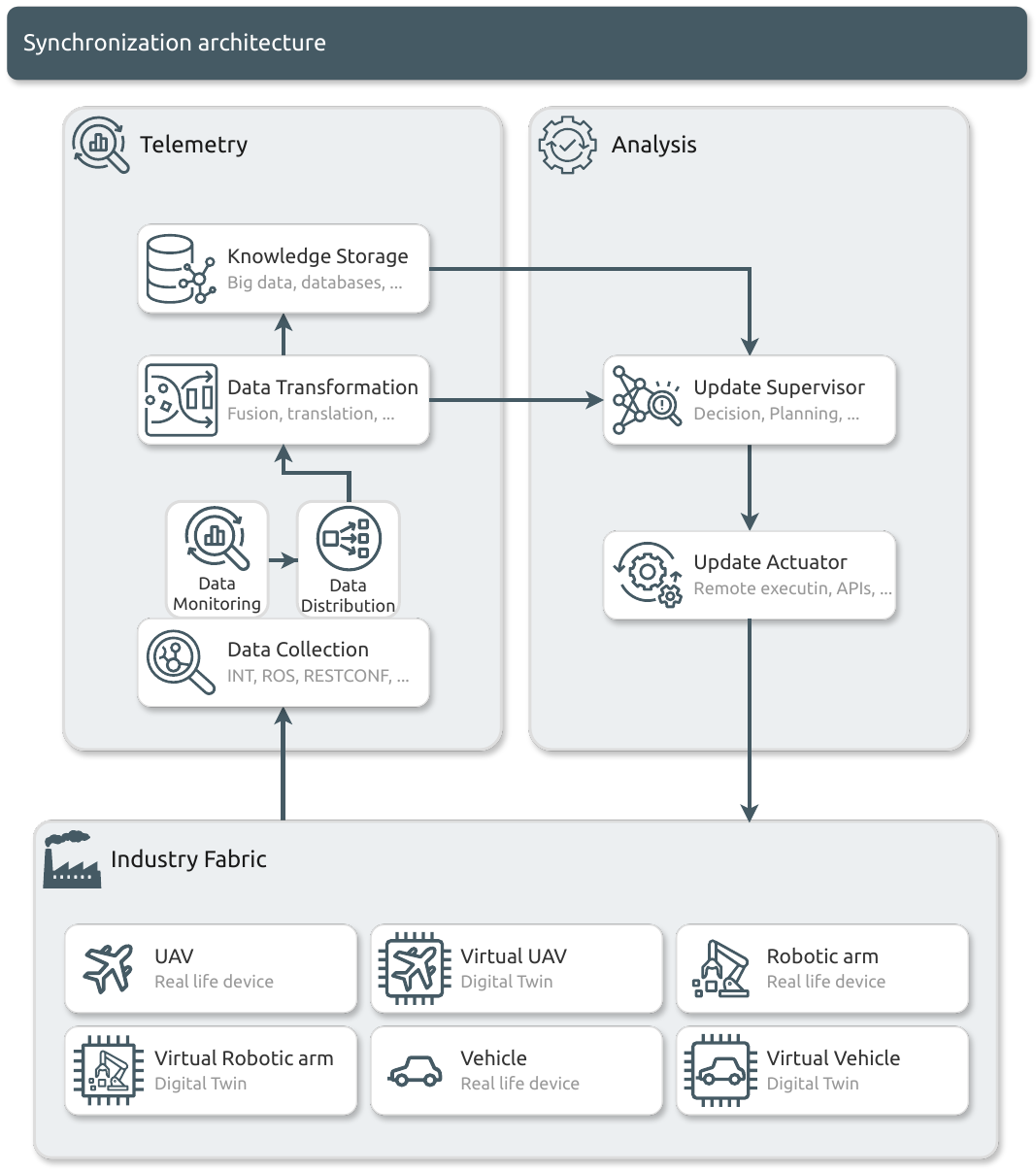}
  \caption{Digital Twin Synchronization Architecture}
  \label{fig:sync-architecture}
\end{figure}

To perform the synchronization of a Digital Twin scenario, there is a need to meet multiple requirements and follow important steps. Articles such as \cite{JIANG2021100196} and \cite{9103025} show that synchronization requires multiple entities, each with its role, such as data collection, data processing, and scenario analysis. Combined, they determine the need for new updates. Analyzing the literature, we see that different works adopt scenario-dependent synchronization methods. However, to the best of our knowledge, no existing contribution specifies a generalized architecture that can be applied across current industrial scenarios. This section describes a synchronization architecture design, the role of each architectural module in the synchronization process, and the way they communicate with each other. An illustration of the proposed architecture is available in Figure~\ref{fig:sync-architecture}. Next, its main modules are introduced. 


\textbf{Telemetry.} At the heart of the synchronization is the Telemetry entity. It represents a mechanism borrowed from network computing that consists of the observability of a network and its services \cite{opentelemetry}. In the context of DT Synchronization, Telemetry consists of collecting, processing, and storing all data possible from both the virtual and real world. It is an extensive part of the synchronization, mainly because it relies on a considerable number of monitoring agents to collect data from each entity in the industry fabric. Also, combining all data objects is an important and challenging step, and filtering, removing, and transforming each device's data is essential in providing clean data for storage and processing. Telemetry consists of three major components, described next, each playing a role in the observability of industry devices.

\textit{Data Collection}, the first Telemetry component, collects data from devices part of the industry fabric. Examples of such physical devices are robots, IoT devices, cameras, routers, and operators. Digital twin devices, on the other hand, are their virtual counterparts. Still, in the \gls{dt} synchronization domain, the industry fabric consists of both physical devices and also digital twin devices. This is because we also need to collect telemetry data from the digital twins since we can have \gls{ai}/\gls{ml} models running algorithms on top of the DT scenario.

\textit{Data Transformation} is the second component of the DT Telemetry entity. It embodies techniques used to transform data from different sources and points in time, facilitating its efficient and safe subsequent processing and storage. Multiple technologies are part of the data transformation component. Among these is Data Fusion \cite{10.1063/1.5031520}. It combines data from multiple sources to create a more comprehensive and representative dataset. Data fusion summarizes data in order to discover patterns and insights that were difficult to observe before, facilitates the detection of events of interest, and estimates device status \cite{DIEZOLIVAN201992}.

Another technology that composes Data Transformation is Knowledge Translation, which transforms the raw telemetry data into \textit{Knowledge Objects}. Based on the work \cite{10154442}, we consider a Knowledge Object as a structure that carries knowledge attributes representing a device or a category of devices. The organization of these attributes leads to less bandwidth utilization and storage usage and requires fewer resources to transmit and store device information \cite{10154442}.

The data collection step provides raw device data regarding different aspects of each device in the industry fabric. For example, a presence sensor can provide data like the presence detection and the sensor's current health status, such as its energy consumption and resource usage. In this example, a single device provides three different types of data that may relate to data from other devices. The Data Transformation component is responsible for organizing and categorizing this type of data so that one can aggregate related data together and still represent the devices and locations they are associated with.

\textit{Knowledge Storage} is the last component of the Telemetry architectural entity. Though part of the synchronization service, it is also shared among other modules. It stores the knowledge objects received from the data transformation process and provides them to the following synchronization entity for analysis and processing. However, information maintained within the {Knowledge Storage} component also includes \gls{ml} and deep learning models.

The {Knowledge Storage} component has a ``sub-component'' referred to as \textit{Data Types} definition. It offers an ontology-like definition based on the type of device, device's information, and category that the data collection is expected to monitor. The definition of data types is a step performed when configuring the synchronization module because it dictates how data is stored and exchanged as part of the synchronization process, but it belongs to the KS module since it relates directly to the data stored in this component.

The digital twin architecture is \textit{data-centric} as stated in \cite{JEREMIAH2024103120}. The main -- if not the only -- data input of the DT must be the telemetry entity since it consists of multiple components organized to ensure proper collection and storage of this data.

\textbf{Analysis.} The next important architectural entity is analysis. It is responsible for performing constant analysis on the data received by Telemetry via the {Knowledge Storage} and {Data Collection} components. It analyzes the data from Telemetry from both worlds and checks if there are updates to perform. If so, it will plan the change requests according to the data supplied, determining the device that needs updating and the possible actions necessary to perform it. Then, it will implement the update commands based on the planned update. It has distributed components, similar to the data collection, that are node-specific and connect directly to the device in the virtual or real world and execute the necessary update. For instance, when the {Analysis} component observes through its processing of Telemetry data that a physical robot position has changed, it identifies the corresponding virtual twin robot and issues appropriate calls to update its position in the virtual world. The result is a synchronization state of both physical and virtual robots.  

The following sections will detail each entity and present a survey of practices and techniques one may leverage to efficiently synchronize digital and real devices properly.

\section{Telemetry}
\label{sec:telemetry}
\begin{figure}[ht]
  \centering
  \includegraphics[width=0.45\textwidth]
  {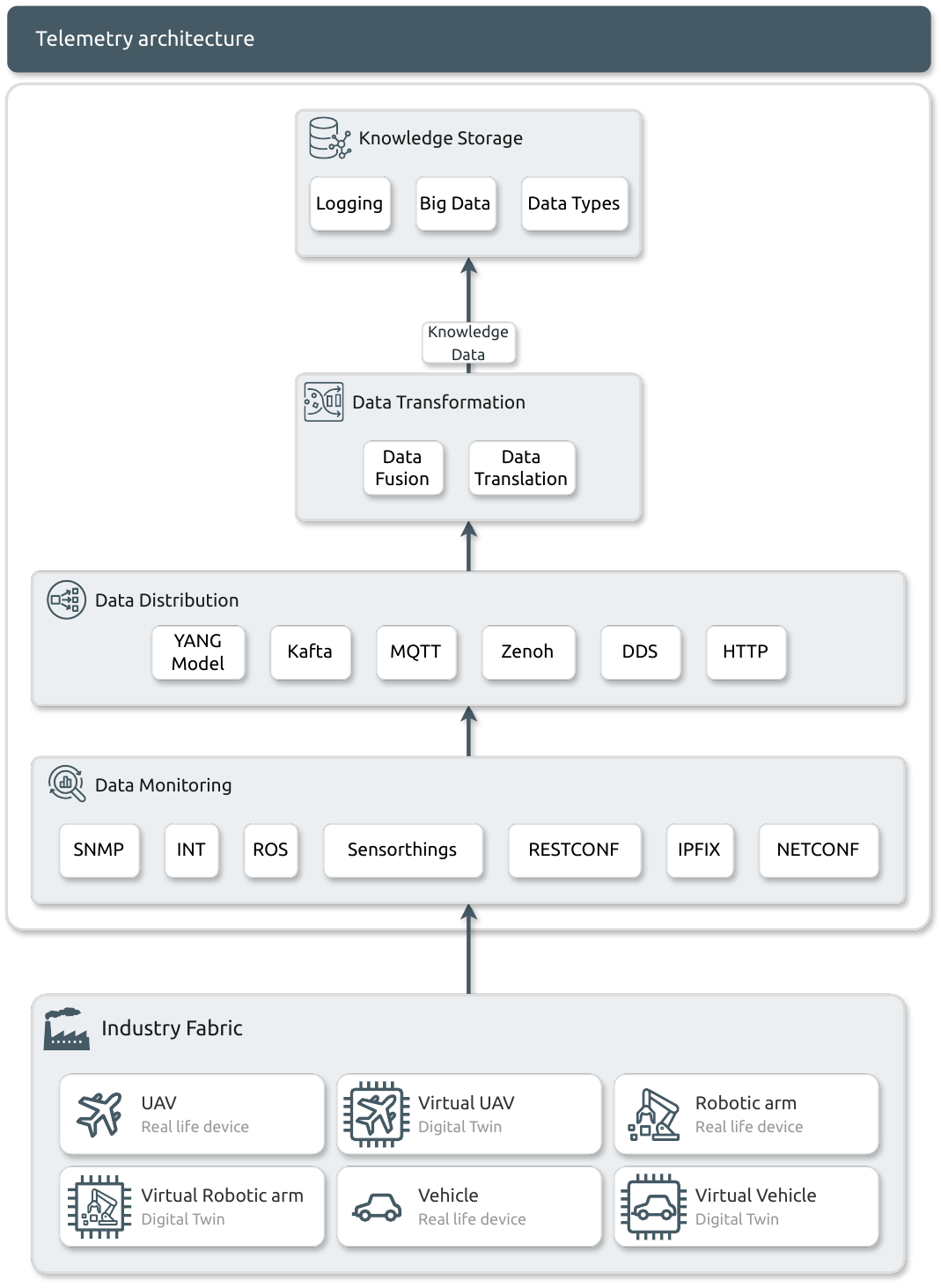}
  \caption{Digital Twin Synchronization Telemetry Architecture}
  \label{fig:telemetry-architecture}
\end{figure}

In this section, we dive deeper into the synchronization's Telemetry entity, detailing its components and discussing the manners to implement it. We divide this section into three subsections, each representing a respective Telemetry component. We also further split Data Collection into three subsections to cover the monitoring particularities of the industry's most significant components: network, robots, and IoT devices. We illustrate the inner details of the telemetry architecture in Figure~\ref{fig:telemetry-architecture}.

\subsection{Data Collection}
\label{sec:data-collection}
Data collection is the most straightforward component of Telemetry. It gathers all possible data that the DT models and synchronization may require. The collection combines a set of technologies, protocols, and tools that interact with the devices to retrieve information, status, and state from these. Next, the collected data is distributed to the relevant {Data Transformation} components. {Data Collection} initiates with \textit{monitoring} industry physical and virtual devices where information is retrieved from. Although it may seem simple to perform, this step can be complex due to the diversity and heterogeneity of devices encountered across the industry domain. A second part of {Data Collection} is the \textit{distribution} of the collected data. After monitoring, one needs to address this data to the appropriate interested parts using supported transmission protocols, which can be manifold depending on the data type and its sources.

\subsubsection{Data Monitoring}
\label{sec:data-monitoring}
Monitoring data is the first step of the synchronization process. Considering the widespread nature of an industry fabric, monitoring consists of a group of distributed agents, each monitoring data from one or more devices of similar features. Different industry domains, such as robotics and telecommunications, have developed specific protocols, information models, and communication standards to perform device monitoring. As a result, this section reviews multiple protocols, middleware solutions, and other technologies used for device monitoring and control, and we discuss previous works that implemented such protocols.

\paragraph{\textbf{Network Telemetry}}

When monitoring network devices for the \glspl{ndt}, we can use the extensive standard network data collection technologies available \cite{zcz-nmrg-digitaltwin-data-collection-03}. Protocols like the Simple Network Management Protocol (SNMP) or Deep Packet Inspection (DPI) tools are candidates for the continuous collection of network information, as well as command-line interfaces (CLI) and system logs for servers. These traditional network monitoring standards were usually performed in a bottom-up approach, where the protocols and technologies deployed to perform monitoring are chosen per device and configured individually. After proper configuration, a lot of unnecessary data must be cleaned to obtain the desired state of each device and the whole network \cite{10.1145/3314212.3314215}.

There is a preference not to use these technologies nowadays for multiple reasons \cite{TAN2021107763} \cite{rfc9232}. Such technologies provide poll-based data collection that hinders \glspl{ndt} Telemetry, which needs continuous updates on the state of network devices. Also, some protocols like SNMP are limited with regard to the type of data they can collect as they cannot be, for example, extended to gain access to new data not previously planned to be collected. 

A data model is also necessary. As we mentioned previously in this paper, structuring the collected data is an essential process since there are multiple steps down in the pipeline that will process and transform this data. Traditional network monitoring techniques also lack the possibility to define data models, which is another drawback for their adoption in this context.

Because of the abovementioned, there is a preference for protocols that refer to the paradigm of ``\textit{Network Telemetry}'' \cite{rfc9232}. As reviewed in \cite{TAN2021107763}, traditional active and passive monitoring gave its way to the emerging {Network Telemetry} concept, where new protocols like RESTCONF or In-band Network Telemetry (INT) reshaped the observability of networks, with more scalable and accurate methods and providing widespread device compatibility \cite{kim2015band}.

{Network Telemetry} reflects the process and set of technologies and protocols leveraged for collecting and delivering network data for outer \textit{management, monitoring, and operation}. The key aspect of {Network Telemetry} is that it is expected to serve autonomous networks, creating network data dedicated to other applications programmed to interpret and act on the received data \cite{rfc9232}. This is crucial to Digital Twin Synchronization since it is a fully automated loop. Also, the programmability of {Network Telemetry} allows full customization regarding which data to collect, the way to model this data, and how to process the data that can be of any necessary size, according to scenario requirements.

{Network Telemetry} is a broad concept, one that cannot be entirely covered by a single technology or protocol. As detailed in \cite{rfc9232}, we need multiple modules to compose a {Network Telemetry} architecture. They include communication protocols, middleware, data types definition, and other technologies. The \textbf{NETCONF} protocol is one such example \cite{rfc6241}. It was specially developed for device configuration management but may serve for {Network Telemetry} purposes since it allows the gathering of device and network-related configuration data. It is based on \gls{rpc}, where a monitored device encodes the data in an XML file and sends it to a monitoring server. DT Synchronization may certainly use the {NETCONF} protocol in order to monitor network device configuration and maintain the configuration of their virtual counterparts updated. Since it is based on \gls{rpc}, there are efficient ways to use NETCONF in a telemetry scenario achieved using, for example, \textbf{YANG-Push} \cite{rfc8641}. YANG-Push extends NETCONF by allowing a publish/subscribe style connection to retrieve telemetry data from network devices, creating a data model for the protocol and a continuous data stream for each device. Similarly, \textbf{RESTCONF} \cite{rfc8040} extends NETCONF to allow data collection based on HTTP methods with a REST Application Programming Interface (API) while also providing backward compatibility with existing NETCONF servers and YANG data types.

Another commonly used protocol for network telemetry is \textbf{\gls{ipfix}} \cite{rfc7011}. This protocol is based on the flow monitoring concept, which consists of exporting every network packet's header to perform analysis and obtain network traffic statistics externally. The fact that we export only the packet header characterizes this as \textit{flow} monitoring instead of packet monitoring, where one would export the entire packet. The flow export in IPFIX consists of a few steps according to \cite{6814316}. It initiates with the definition of information elements, which are entries in a table to a combination of information extracted from headers, such as destination TCP port or source IP address. It is important to note that these information elements can be from any TCP/IP standard protocol. The flow caches are where the information elements are stored as cache tables and organized as flow keys. The flow keys organize the flows to identify new packets in a flow. After the packets are identified as flows, sampling, and filtering are necessary to simplify the flows and optimize packet exporting. Next, the IPFIX message can be created with information about the flow or set of flows present in the message and the flow itself. Once the IPFIX message is created, a transport protocol is selected, and the message can be exported. Flow monitoring is another key aspect of network telemetry for DT Synchronization. Since the {Network Digital Twin} can be a model inside a simulation/emulation environment, we need the base model data to initiate the DT. Part of this data must also be about how the network traffic flows behave, and this data is constantly changing. Monitoring the network flow is essential to keep modeling the NDT, and IPFIX is an adequate candidate to allow the NDT model to remain updated with the real world \cite{10154442}. 

\textbf{In-Band Network Telemetry} is an abstract concept that combines multiple underlying technologies. It is also known as Data Plane Telemetry or In-line Telemetry. The concept proposes the collection of metrics through the injection of network device information inside the real network packets themselves, hence the use of the term {In-Band} as opposed to {Out-of-Band} telemetry using external protocols \cite{TAN2021107763}. Each hop on the network inserts specific telemetry data, and the last one delivers the combined data to the telemetry server. There are multiple in-band protocols to choose from; among these is the INT, a literal abbreviation that means In-band Network Telemetry. INT operates by sending telemetry instructions on the network. A node that receives this instruction starts collecting and inserts telemetry data inside the network packets. INT collection works directly on the data plane without involving control plane management. Another protocol implementation is Alternate Marking-Performance Measurement (AM-PM), which estimates delay and packet loss on the network between two endpoints. Each packet is marked with a bit on its header that indicates the fragmentation of packets on smaller packets on a single flow, which is used to synchronize the measurement of network performance metrics. The paper \cite{PEZAROS20103246} shows an INT scenario that uses the IPv6 protocol and its extension header fields to attach telemetry data. The protocol choice allows a flexible implementation of the type of information needed to collect while it maintains backward compatibility with devices that are unaware of the protocol extension. Because of this, INT is considered an appropriate candidate to compose Digital Twin Data Collection, especially for maintaining network state in the digital world.

\paragraph{\textbf{ROS}}
Other technologies are available when monitoring robots in an industrial scenario for the DT. The monitoring task is the same: collect the necessary data for the DT synchronization and forward it to the distribution node. However, the protocols used to perform this may differ from those used in the network monitoring. Lately, robots have been using Robot Operating System (ROS) \cite{ros} to enable integration with outside networks and controllers \cite{10.22260/ISARC2020/0205} \cite{exploiting-cell} \cite{9893160}. ROS is not a full operating system besides its name. It rather runs on top of an existing operating system and includes tools and technologies that allow the communication of multiple robots and workstations to interact with the robots. It uses a standard publish/subscribe communication pattern to establish communication between its nodes and sets data types to maintain a structured communication of the information between them \cite{quigley2009ros}.

ROS combines software libraries and tools designed to create and manage robot applications and communications. It enables the insertion of these tools inside a robot and provides the proper ways to interact with it, control its movements, extract status information, and automate its trajectory. With regard to DT Telemetry, ROS can be an important player useful to interact with the robots and collect telemetry data related to the robot's joint movements or geographic location, as well as monitoring its current status, including information on CPU or memory usage \cite{exploiting-cell}.

Additionally, ROS can be used to transmit these data to other robots in order to synchronize them, where a variety of protocols can be used for communication between modules \cite{diachenko2022industrial} \cite{singh2024unity}. The work \cite{singh2024unity} uses a ROS framework to send and receive data from Unity using TCP, acting as an intermediary in the data exchange. Works like Wang et al. \cite{wang2023bidirectional} and Liang et al. \cite{10.22260/ISARC2020/0205} created a bridge for bi-directional data monitoring, with \cite{wang2023bidirectional} using a ROS framework called ROSBridge to transform the data in JSON format and exchange it using common communication protocols like WebSocket. Similarly, \cite{10.22260/ISARC2020/0205} creates its own method to transform the data into a Python string and exchange the data using middleware communication protocols.

However, creating a custom programming interface to interact with a robot and other modeling technologies is also possible, like the works in \cite{tahemaa2019digital} and \cite{SCHROEDER201612}. These contributions created specific programs and scripts to collect the location and movement status of a robot. The work uses its own interfaces to send the data to the update analysis, where DT synchronization takes place.

\paragraph{\textbf{SensorThings API}}
When referring to IoT devices as part of the digital twin, other constraints become relevant. Usually, one cannot run full monitoring applications inside these devices, which are often energy or resource-limited. Commonly, we have physical monitors to perform data collection \cite{app10186519}. On the other hand, these monitoring devices have their own specific constraints, such as energy consumption or geo-location. Because of this, specific protocols are available that aim to address these constraints. A notable one is SensorThings API \cite{sensorthings}. It provides a framework that connects IoT devices and transmits data over the web, allowing orchestration of ``sensing'' and ``actuating'' over the IoT devices. Actuating refers to the operation of actuators in the network that can perform updates on the devices they connect to. Sensing is data collection regarding the state of IoT devices and other related metrics. Communication is performed using REST-based services and the MQTT publish/subscribe paradigm. It also supports data structuring with JavaScript Object Notation (JSON) serialization.

The {SensorThings API} framework can be useful for DT Synchronization since it provides standard and mature ways to perform data collection of specific IoT devices.

\subsubsection{Data Distribution}
\label{sec:data-distribution}
The second part of data collection is its distribution. As the data's size may sometimes be significant, it can be challenging to transport. As a result, a lightweight distribution methodology may be adopted. Some of the mechanisms that optimize {Data Distribution} are presented next. 

\paragraph{\textbf{Modeling}}
First, data modeling is essential and can rely on schemas such as the YANG data model \cite{rfc8639} \cite{rfc7950}. YANG enables the modeling of state and configuration data and notifications for management protocols, which are needed by the synchronization service.

\paragraph{\textbf{Transmission}}
Second, there is a need for an efficient transmission protocol. Standard poll-based data collection, usually found in the more traditional network management protocols such as SNMP, can hinder data distribution performance and interfere more than needed through its traffic overhead \cite{rfc9232} \cite{8088251}. Publishing the collected data in a stream and subscribing to this stream can reduce the latency for change detection \cite{rfc9232}. Message Oriented Middleware is helpful since it can connect multiple devices that will then send collected information from the monitoring agents to data transformation components in a way that prioritizes bandwidth efficiency and low energy consumption \cite{9559032}. There are multiple publish/subscribe middleware solutions one may choose from, such as \gls{MQTT}, \gls{AMQP}, Zenoh, Kafka, \gls{DDS}, and others. Parameters like available resources, number of devices on the scene, or type of messages sent are essential to determine an appropriate implementation through careful evaluation \cite{8088251} \cite{9559032} \cite{10.1145/3409334.3452067} \cite{s19194217}.

\subsection{Data Transformation}
\label{sec:data-transformation}
{Data Transformation} is an intermediate stage taking place between data collection and storage \cite{exploiting-cell}. Digital twins tend to generate massive volumes of data, ranging from real-time device states to human safety warnings, for example. Such data needs to be processed by the {Analysis} component as well as possibly other data-driven modules such as those implementing deep learning algorithms as part of their application. As a result, there is a need to transform this massive amount of data to facilitate its processing and storing, making it noise-free, accurate, and reliable, all without information loss \cite{CSI}. A wide range of technologies transforms data, including data fusion, modeling, cleaning, evolution \cite{8477101}, integration, and translation tasks \cite{10154442}.

Data fusion is a concept stemming from the {Information Fusion} field. It combines data from multiple sources to create a more comprehensive and valuable dataset than what one would achieve when using individual data sources in an isolated manner. Data fusion combines the collected data so that data storage can be efficient, noise-free, accurate, and reliable without losing information \cite{8477101}. For example, data fusion, when applied in an industry scenario, can benefit a damage detection application, reducing inference time and ambiguity while maintaining high accuracy \cite{HASSANI2024102136}. Recall that DT synchronization is responsible for collecting data from the industry fabric and that this data is the base for synchronizing both the real and virtual worlds. However, this data must often be properly stored in a time series database and may also serve as the dataset for the deep learning and artificial intelligence models executing as part of the Digital Twin \cite{8477101} \cite{10.1063/1.5031520}. Because of this, data fusion becomes an even more critical step in data transformation because it integrates and processes the collected data so that synchronization and predictive/diagnoses models can be easier to perform.

In addition to {Data Fusion}, {Data Translation} is crucial to boost transmission, storage, and synchronization analysis. This step translates the collected data into knowledge data, hence optimizing storage resources. This translation may be performed in multiple ways depending on whether the data to translate is the raw telemetry data collected by the monitoring or the result of already pre-processed data from data fusion. The scenario at hand will dictate how and what will need to be translated. The work \cite{10154442} proposes ``Telemetry Knowledge Objects'' (TKOs), which is a concentration of information obtained by the network node, converted to the standard model for data interchange on the Web, namely, the {Resource Description Framework} (RDF). RDF has features that facilitate data merging even if the underlying schemas differ \cite{klyne2004resource}. The term knowledge objects is effective in DT synchronization because it focuses on extracting only the essential information from the collected data, discarding any additional information that usually comes with standard Telemetry. Knowledge objects are lighter to transmit, and the amount of \textit{useful} information remains higher and precise, which enables higher processing power to the management system and less bandwidth occupation.

\subsection{Knowledge Storage}
\label{sec:knowledge-storage}
The last telemetry component is also a dual component, as it contributes both to synchronization and is needed by AI ML/DL models of the Digital Twin. We refer to it as Knowledge Storage (KS) since it stores \textit{knowledge} instead of raw data \cite{10154442}. Knowledge Storage (KS) is seen as a type of Data Warehouse designed to centralize and organize data in different formats -- including processed information and insights generated by DL/ML/AI models. It stores historical data and facilitates real-time querying and analysis, which is essential for the continuous operation and optimization of the Digital Twin.

KS could leverage robust Data Warehouse and Data Lake technologies to manage diverse data types. Platforms like Apache Hadoop \cite{shvachko2010hadoop} and Apache Spark \cite{zaharia2016apache} are prominent for handling large volumes of structured and unstructured data, supporting advanced analytics and machine learning workflows. Cloud-based solutions such as Amazon S3, Azure Data Lake Storage, and Google Cloud Storage provide scalable and cost-effective storage options for Data Lakes \cite{hashem2015rise, sandhu2021big}. In addition, technologies like InfluxDB and Prometheus offer optimized storage solutions for time series data common in continuous monitoring systems \cite{nasar2019suitability, turnbull2018monitoring, Chakraborty2021}. They efficiently handle data that varies over time, allowing the capture of high-frequency metrics and enabling real-time querying.

KS can be configured to be accessed in different ways. REST APIs are a common and efficient approach, allowing data to be consumed and updated by other applications in a scalable manner. Alternatives are also viable. gRPC, for example, is a technology that offers fast and efficient communication between services, which is ideal for high-performance scenarios. In addition, messaging protocols such as MQTT can be used for real-time data transmission, especially in IoT environments, ensuring that updates are published in a lightweight and quick manner. InfluxDB and Prometheus provide RESTful interfaces and support specific queries via their own query protocols, such as PromQL, allowing flexible integration with other systems. These capabilities ensure that DL/ML models can consume the latest data and publish their predictions directly to KS through different means. This continuous communication between the models and the central system is essential for updating the Digital Twin.

In addition to its role as a data repository, the KS stores information and results generated by AI models. When these models consume KS data to predict events or perform analyses, their outputs, such as predictions or alerts about the need for adjustments to the Digital Twin, are recorded back in the KS. Analytics engines then consume this output, forming a continuous feedback loop. This approach ensures that real-time operations reflect AI-driven decisions, enabling constant system optimization.

This data input and output cycle creates a truly data-centric system, where data not only feeds models but also returns to the repository as output that is continually used to optimize and adjust the Digital Twin. In this way, Knowledge Storage can be seen as a dynamic data storage and analytics platform, where the continuous flow of information between sensors, models, and analytics ensures that the Digital Twin is always aligned with the physical world, ready to predict and react to changes in real time.

Candidate technologies for the support of {Knowledge Storage} are rooted in the Big Data field \cite{exploiting-cell} \cite{DIHAN2024e26503} \cite{9103025}. They include mainly four types of solutions that depend on their underlying data models: key-value models (with storage tools such as Redis and Amazon DynamoDB), column-oriented data models (adopted by Apache Cassandra and HBase), document-oriented data models (offered by database technologies such as MongoDB and Couchbase), and graph-based data models (implemented using Neo4j and Amazon Neptune). Among these, the most suitable data models for big data are key-value and graph-based knowledge storage approaches due to their efficiency in data retrieval and capability for modeling complex relationships \cite{Siddiqa2017}.

For each of the four approaches, data is accessed differently. Key-value systems provide fast access through unique keys, making these ideal for low-latency queries and applications requiring real-time or "near-real" responses. Graph databases allow queries based on relationships between entities, such as in Neo4j, and are optimized to model complex interactions, commonly encountered in Digital Twins, which usually involve other interconnected entities -- such as sensors, devices, processes, and people \cite{RAMONELL2023105109}. On the other hand, column-oriented systems such as Cassandra are practical for large volumes of structured data, while document-oriented systems (e.g., MongoDB) are helpful for semi-structured data.

\subsubsection{Data Types}
When monitoring the industry's device, one must know what it seeks to collect. Based on this information, one defines a data structure that will specify the type of data that will be collected, the sources of these data, and the manner in which it will interact in data transformation and storage. Data Type definition is, therefore, an essential step for data collection \cite{DIHAN2024e26503}.

\section{Analysis}
\label{sec:analysis}
\begin{figure}[ht]
  \centering
  \includegraphics[width=0.45\textwidth]
  {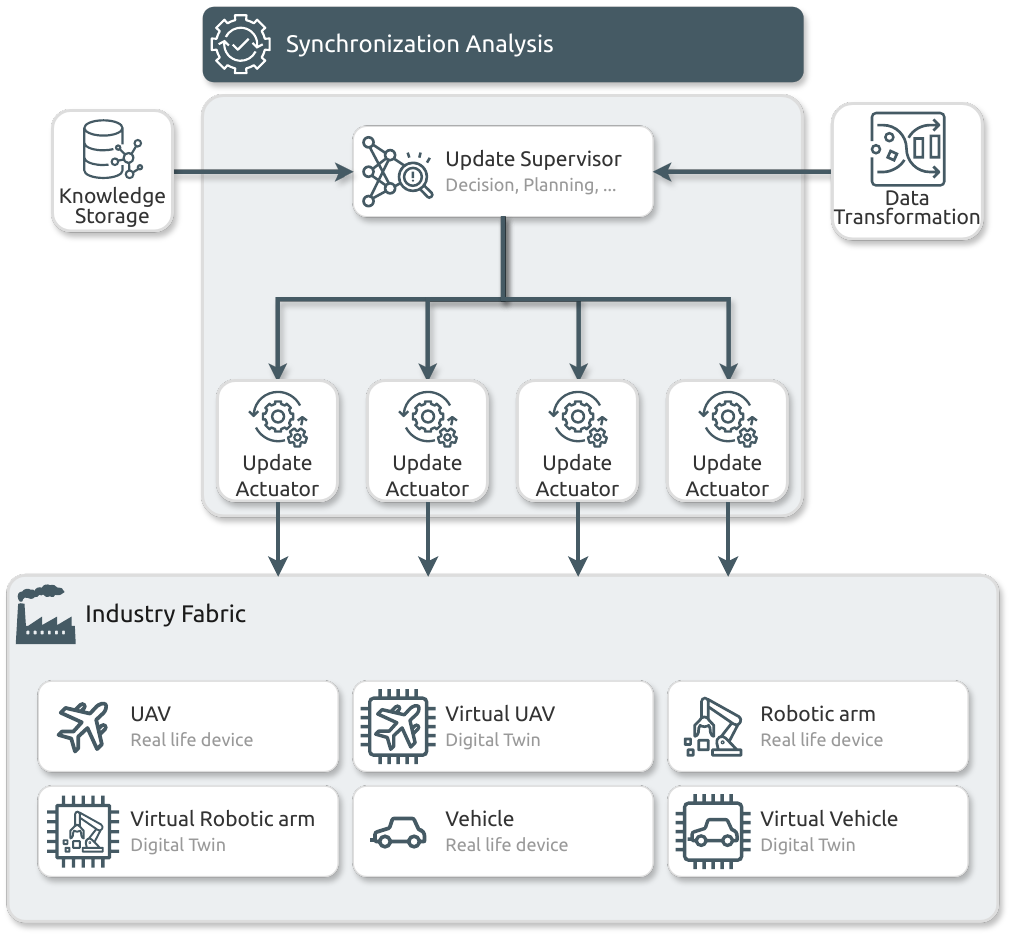}
  \caption{Digital Twin Synchronization Analysis Architecture}
  \label{fig:analysis-architecture}
\end{figure}

This section further details the analysis entity as part of the synchronization service. Each subsection of this section represents a component of the {Analysis} architectural entity as illustrated in Figure~\ref{fig:analysis-architecture}.

\subsection{Update Supervisor}
\label{sec:update-supervisor}
The update supervisor is one of the most critical components of synchronization. It is responsible for verifying whether an update is necessary for every device in the industry fabric. It may be a distributed supervisor, where each agent is responsible for verifying a specific or a group of device(s) or a single object responsible for overlooking the updates necessary to the system as a whole. The choice for distribution depends on the problem size. First, the supervisor subscribes to the data collection queue on the telemetry side. Every time a new status update is available, the supervisor receives it.

Then, it compares the status of both the virtual and real world to detect if there is a need for updates. Here, the supervisor should also confer the {Knowledge Storage} to verify whether the update is valid. For example, a previous update attempt may have failed, and the knowledge storage should gain access to this information. A detailed logging of each update attempt should be performed, containing details of the device, the nature of the update, and the reasons for failure, for example \cite{Chu2018Failure, Wetterneck2006Using}. This data can be used as failure history to identify recurring patterns that can increase the success rate of future updates. The supervisor is responsible for verifying if there is a failed attempt to update in order to ensure the retrial of a failed attempt.

In addition, the supervisor should check the knowledge storage. As previously mentioned, the KS stores every data item the monitors collect, which may be consumed by \gls{ml} models used in DT applications. If a model's output relates to synchronization, it should publish an output for an update intention inside the storage. Examples of models that relate to synchronization are collision detection between robots and humans or industrial network congestion. In cases like this, machine learning models can predict danger and issue an update request to prevent such events. This update request converts into an industry fabric update and, therefore, a synchronization task. The models should post this update on the {Knowledge Storage}, where the update supervisor can fetch and verify it. Then, it should perform a reverse verification, where it checks with the up-to-date data if the scenario still needs to be updated with the model's update request or if the scenario has already been updated externally. The update decision is a critical step in the synchronization process. As reviewed in \cite{10015424}, synchronizing the \gls{dt} at a state-dependent policy reduces the number of actual synchronization, even though it creates a higher number of synchronization possibilities and observations.

After detecting the need for an update, the supervisor plans the update, checking for anything necessary to perform the update correctly. For example, there may be a need to check the device API to ensure connectivity or to ``activate it''. Also, one device may have multiple update options; for example, a mobile robot can have moving parts, and the update may refer to its location or its joints. A plan is drawn to categorize and ensure the proper execution of the update. When the plan is ready, the supervisor publishes the updated intention in the appropriate queue for the responsible actuator.

\subsection{Update Actuator}
\label{sec:update-actuator}
The Update Actuator is responsible for effectively updating the required devices. It is a distributed component, similar to {Data Collection} in {Telemetry}. There should be one actuator for each device or at least a group of similar devices, as the actuator sends device-specific updates deeply related to the device's own API or system configuration.

Once the supervisor plans the update and publishes it to the actuator queue, the actuator receives it and implements it. The update intention that the actuator receives assumes that the device is working and ready to receive updates and that the update is valid and possible to implement. The actuator should not perform any verification, only the update itself.

\section{Conclusions}
\label{sec:conclusion}

This study addressed the importance of Digital Twin technology in the context of Industry 4.0, highlighting the need for effective synchronization between the physical and virtual worlds. The literature review indicated significant advances but also revealed critical challenges related to device heterogeneity and data complexity.

This survey contributes by proposing a unified architecture for Digital Twin synchronization. The implementation of this architecture aims to standardize processes and promote interoperability between systems, facilitating the adoption of Digital Twins in diverse industrial sectors. Thus, this work not only identifies gaps in the existing literature but also suggests practical solutions to improve the effectiveness and efficiency of DT systems, aligning with the objectives of Industry 4.0.

In the future, we plan to evaluate the overhead of the synchronization service in a working DT environment. Depending on the scenario at hand, one expects that complete synchronization may be costly to achieve in terms of computing, storage, and communication resources. We will look at finding suitable tradeoffs between this overhead and the accuracy of the synchronization service. This may be achieved, for example, by limiting the frequency of synchronization events and differentiating among them through some processing priority mechanism.

\section{Acknowledgments}
\label{sec:Acknowledgments}

Fundação de Amparo à Ciência e Tecnologia do Estado de Pernambuco (FACEPE), Coordenação de Aperfeiçoamento de Pessoal de Nível Superior (CAPES), Conselho Nacional de Desenvolvimento Científico e Tecnológico (CNPq), supported this work.

\bibliographystyle{elsarticle-num-names}
\bibliography{references}

\end{document}